\documentclass[preprint,a4paper,onecolumn,12pt]{revtex4}%
\usepackage{amsfonts}
\usepackage{amsmath}
\usepackage{amssymb}
\usepackage{graphicx}%
\providecommand{\U}[1]{\protect\rule{.1in}{.1in}}

\begin{document}
\title{Unitarity of the time-evolution and observability of non-Hermitian
Hamiltonians for time-dependent Dyson maps}
\author{F. S. Luiz$^{1}$, M. A.de Ponte$^{2}$, and M. H. Y. Moussa$^{1,3}$}
\affiliation{$^{1}$Instituto de F\'{\i}sica de S\~{a}o Carlos, Universidade de S\~{a}o
Paulo, Caixa Postal 369, 13560-970, S\~{a}o Carlos, S\~{a}o Paulo, Brazil}
\affiliation{$^{2}$Universidade Estadual Paulista, Campus Experimental de Itapeva,
18409-010, Itapeva, S\~{a}o Paulo, Brazil}
\affiliation{$^{3}$Centre for Mathematical Science, City University, Northampton Square,
London EC1V 0HB, UK}

\begin{abstract}
Here we present an strategy for the derivation of a time-dependent Dyson map which ensures simultaneously the unitarity of the time evolution and the observability of a quasi-Hermitian Hamiltonian. The time-dependent Dyson map is derived through a constructed Schr\"{o}dinger-like equation governed by the non-Hermitian Hamiltonian itself; despite its time-dependence our scheme ensures the time-independence of the metric operator, a necessary condition for the observability of the quasi-Hermitian Hamiltonian. As an illustrative example we consider a driven Harmonic oscillator described by a time-dependent non-Hermitian Hamiltonian. After computing the Dyson map and demonstrating the time-independence of the associated metric operator, we successfully derive an eigenvalue equation for this time-dependent Hamiltonian which enable us to analyze the $\mathcal{PT}$-symmetry breaking process.
\end{abstract}

\maketitle

\textit{Introduction}. Since the work by Bender and Boettcher \cite{BB} and further developments by Mostafazadeh \cite{Mostafazadeh}, quasi-Hermitian $\mathcal{PT}$-symmetric Hamiltonians have been extensively studied. While in the former reference it was suggested that Hamiltonians invariant under space-time reflection symmetry ($\mathcal{PT}$-symmetry) can have real spectra, in the latter the notion of quasi-Hermiticity was introduced, establishing the grounds for treating non-Hermitian $\mathcal{PT}$-symmetric Hamiltonians using time-independent metric operators \cite{Grounds}. Aiming to extend the scope of Hermitian quantum mechanics, the steps towards the deepening of our understanding of non-Hermitian systems has since been taken in virtually all fields of physics \cite{Grounds}. More recently, $\mathcal{PT}$-symmetry (PTS) and PTS breaking has been investigated in a variety of systems, such as waveguides \cite{WG}, optical lattices \cite{OL} and optomechanics \cite{OM}. Moreover, a variety of phenomena such as disorder \cite{Disorder}, localization \cite{Localization}, chaos \cite{Chaos} and solitons \cite{Solitons} have been investigated within $\mathcal{PT} $-symmetric systems.

Despite the overall consensus on handling quasi-Hermitian Hamiltonians through time-independent metric operators \cite{Grounds}, controversies emerged regarding the generalization to time-dependent (TD) metric operators
\cite{Mostafa,Controversy}. Although it has been demonstrated that a TD metric operator can not ensure the unitarity of the time-evolution simultaneously with the observability of the Hamiltonian \cite{Mostafa}, some authors have disputed this claim \cite{Controversy}, failing however to ensure the unitarity of time evolution by insisting on the observability of the Hamiltonian. A contribution has been recently presented in Ref. \cite{Fring} for dealing with TD metric operators which, although in agreement with Ref. \cite{Mostafa}, goes a step beyond; It has been demonstrated that a TD Dyson equation and a TD quasi-Hermiticity relation (as first introduced in \cite{Fring}) can be solved consistently at the cost of rendering the non-Hermitian Hamiltonian to be a nonobservable operator, \textit{but showing that any other observable in the non-Hermitian system is derived in complete analogy with the time-independent scenario}. Non-trivial solutions to the TD Dyson equation and the proposed TD quasi-Hermiticity relation have been presented, starting with a non-Hermitian linearly driven harmonic oscillator and a spin chain \cite{Fring}, and then going to the TD Swanson model \cite{Swanson}.

In the present contribution, however, we present a strategy to go even beyond Ref. \cite{Fring}, enabling us to account for the unitarity of the time-evolution simultaneously with the observability of a non-Hermitian Hamiltonian for a TD Dyson map. To this end a Schr\"{o}dinger-like equation is constructed from which we derive the TD Dyson map from the TD quasi-Hermitian Hamiltonian itself. The key point here is that, despite the time-dependence of the Dyson map our scheme remarkably ensures a time-independent metric operator, a necessary condition for the observability of a quasi-Hermitian Hamiltonian. The distinction here established between the time-dependence of the Dyson map and that of the metric operator is therefore central to the current development. Thus, \textit{although we are in agreement with the main premise in Refs. \cite{Mostafa,Fring}, that a time-independent metric operator is needed for assuring the unitarity of the time evolution simultaneously with the observability of the quasi-Hermitian Hamiltonian, here a TD Dyson map is
considered, and this is an important point since for a TD non-Hermitian Hamiltonian, a time-independent Dyson map is a rather restrictive choice.}

Our Schr\"{o}dinger-like equation applies, however, to a more general scenario than the one for which it was constructed; apart from the TD non-Hermitian Hamiltonians, it also applies to time independent non-Hermitian Hamiltonians, in the latter case recovering exactly the standard procedure for handling non-Hermitian quantum mechanics as we show below. It also helps with unitary transformations within Hermitian quantum mechanics, providing us with the transformation operator from the Hamiltonian itself. As an illustration of our method we revisit the non-Hermitian linearly driven harmonic oscillator, deriving the TD Dyson map from the constructed Schr\"{o}dinger-like equation and showing the time-independence of the associated metric operator. Finally, after deriving an eigenvalue equation for a TD non-Hermitian system, we analyze the $\mathcal{PT}$-symmetry breaking process.

\textit{A Schr\"{o}dinger-like equation for the evolution of the TD Dyson map}. Starting with a brief review of the developments in Ref. \cite{Fring}, we consider a non-Hermitian TD Hamiltonian $H(t)$ associated with the Schr\"{o}dinger equation $i\partial_{t}\left\vert \psi(t)\right\rangle =H(t)\left\vert \psi(t)\right\rangle $. A TD Dyson map $\eta(t)$ thus leads to the TD Dyson relation, i.e., the transformed Hamiltonian
\begin{equation}
h(t)=\eta(t)H(t)\eta^{-1}(t)+i\left[  \partial_{t}\eta(t)\right]  \eta^{-1}(t)\text{,} \label{eq1}%
\end{equation}
which generates the evolution of the equation $i\partial_{t}\left\vert
\phi(t)\right\rangle =h(t)\left\vert \phi(t)\right\rangle $, where $\left\vert
\phi(t)\right\rangle =\eta(t)\left\vert \psi(t)\right\rangle $. Due to the
gauge-like term in Eq. (\ref{eq1}) the non-Hermitian $H(t)$ and its Hermitian
counterpart $h(t)$ are no longer related by means of a similarity
transformation, resulting in that $H(t)$ is not a self-adjoint operator and,
therefore, not observable. The Hermiticity of $h(t)$ leads however, as
referred to in Ref. \cite{Fring}, to the \textit{TD quasi-Hermiticity
relation}
\begin{equation}
H^{\dag}(t)\rho(t)-\rho(t)H(t)=i\partial_{t}\rho(t),\qquad\rho(t)=\eta
^{\dagger}(t)\eta(t)\text{,} \label{eq2}%
\end{equation}
which replaces the usual relation $H^{\dag}\rho=\rho H$ for a time-independent
metric. Assuming $\rho(t)$ to be a positive-definite TD metric operator, it is
straightforward to verify that\textbf{ }the generalized Eq. (\ref{eq2}) leads
to the expected relation between the TD probability densities in the Hermitian
and non-Hermitian systems, given by
\begin{equation}
\left\langle \psi(t)\left\vert \tilde{\psi}(t)\right.  \right\rangle
_{\rho(t)}=\left\langle \psi(t)\left\vert \rho(t)\right\vert \tilde{\psi
}(t)\right\rangle =\left\langle \phi(t)\left\vert \tilde{\phi}(t)\right.
\right\rangle \text{.} \label{eq3}%
\end{equation}
From the above observation one concludes, as in Ref. \cite{Fring}, that even
for TD Dyson maps, any observable $o(t)$ in the Hermitian system possesses a
counterpart $O(t)$ in the non-Hermitian one ---except for the non-Hermitian
Hamiltonian itself--- given by%

\begin{equation}
O(t)=\eta^{-1}(t)o(t)\eta(t)\text{,} \label{eq4}%
\end{equation}
in complete analogy for time-independent Dyson maps.

Our strategy to restore a similarity transformation from Eq. (\ref{eq1}) and,
consequently, to restore the observability of $H(t)$ ---thus going beyond Ref.
\cite{Fring}---, is to impose the gauge-like term $i\left[  \partial_{t}%
\eta(t)\right]  \eta^{-1}(t)$ equal to to the operator $\eta(t)H(t)\eta
^{-1}(t)$, thus leading to the Schr\"{o}dinger-like equation%
\begin{equation}
i\partial_{t}\eta(t)=\eta\left(  t\right)  H\left(  t\right)  \text{,}%
\label{eq5}%
\end{equation}
which enable us to compute the TD Dyson map from the non-Hermitian $H(t)$
itself. The Eq. (\ref{eq5}), which is similar to the Schr\"{o}dinger equation
written in the dual Hilbert space, ensures the similarity transformation%
\begin{equation}
h(t)=2\eta(t)H(t)\eta^{-1}(t),\label{eq}%
\end{equation}
and by demanding $h(t)$ to be Hermitian, we derive the quasi-Hermiticity
relation%
\begin{equation}
H^{\dag}(t)\rho(t)=\rho(t)H(t),\label{Eq5}%
\end{equation}
which, together with Eq. (\ref{eq2}), implies, despite the time-dependence of
the Dyson map, the time-independence of the metric operator, $\rho
(t)=\rho(t_{0})$, a necessary condition for the observability of $H(t)$, the
time-dependent similarity transformation (\ref{eq}) being the necessary and
sufficient condition the observability of $H(t)$. The quasi-Hermiticity
relation (\ref{Eq5}) helps us to define the initial condition $\eta\left(
t_{0}\right)  $ for the exact solution of Eq. (\ref{eq5}), given by%
\begin{equation}
\eta\left(  t\right)  =\eta\left(  t_{0}\right)  T\exp\left[  -i\int_{t_{0}%
}^{t}d\tau H\left(  \tau\right)  \right]  ,\label{eq6}%
\end{equation}
where $T$ denotes the time ordering operator. Except for the initial condition
$\eta\left(  t_{0}\right)  $, $\eta\left(  t\right)  $ is a determinist
operator following from the non-Hermitian $H(t)$. In short, under the
constructed Schr\"{o}dinger-like equation (\ref{eq5}) the TD Dyson equation
(\ref{eq1}) and quasi-Hermiticity relation (\ref{eq2}) reduce to their
simplified forms in Eqs. (\ref{eq}) and (\ref{eq5}) which ensures the
unitarity of the time evolution governed by $H(t)$ simultaneously with the
observability of such a non-Hermitian Hamiltonian. Finally, with the unitarity
of the time-evolution assuming the form of Eq. (\ref{eq3}), the matrix
elements of the observables in Eq. (\ref{eq4}) becomes%
\begin{equation}
\left\langle \psi(t)\right\vert O(t)\left\vert \tilde{\psi}(t)\right\rangle
_{\rho(t)}=\left\langle \phi(t)\right\vert o(t)\left\vert \tilde{\phi
}(t)\right\rangle \text{.}\label{Eq7}%
\end{equation}

In Supplementary Material (SM) we demonstrate from Eq. (\ref{eq6}) and using
(\ref{Eq5}), that $\rho(t)=\eta^{\dag}\left(  t\right)  \eta\left(  t\right)
=\eta^{\dag}\left(  t_{0}\right)  \eta\left(  t_{0}\right)  $, with the
initial value $\eta\left(  t_{0}\right)  $ following from the parameters
defining $H(t)$.

\textit{An illustrative example}. In order to illustrate the method proposed
above, we consider a Hamiltonian of the form
\begin{equation}
H\left(  t\right)  =H_{0}\left(  t\right)  +\kappa V\left(  t\right)
,\qquad\left[  H_{0}\left(  t\right)  ,H_{0}\left(  t^{\prime}\right)
\right]  =0, \label{eq7}%
\end{equation}
where $H_{0}\left(  t\right)  $ stands for the usual free Hamiltonian and
$\kappa V\left(  t\right)  $ stands for a non-Hermitian interaction with a
real dimensionless strength $\kappa$, to be considered as a perturbation
parameter. The TD Dyson map coming from Eqs. (\ref{eq6}) and (\ref{eq7}) is
thus given by%
\begin{equation}
\eta\left(  t\right)  =\eta\left(  t_{0}\right)  \left[  T\exp\left(
-i\kappa\int_{t_{0}}^{t}d\tau\tilde{V}\left(  \tau\right)  \right)  \right]
\exp\left(  -i\int_{t_{0}}^{t}d\tau H_{0}\left(  \tau\right)  \right)  ,
\label{eq8}%
\end{equation}
where $\tilde{V}\left(  t\right)  =\exp\left(  -i\int_{t_{0}}^{t}d\tau
H_{0}\left(  \tau\right)  \right)  V\left(  t\right)  \exp\left(  i\int
_{t_{0}}^{t}d\tau H_{0}\left(  \tau\right)  \right)  $. For a TD harmonic
oscillator under a TD non-Hermitian linear amplification, $H\left(  t\right)
$ is given by
\begin{equation}
H_{0}\left(  t\right)  =\omega\left(  t\right)  a^{\dagger}a\text{,}\qquad
V\left(  t\right)  =\alpha\left(  t\right)  a+\beta\left(  t\right)
a^{\dagger}\text{,} \label{eq9}%
\end{equation}
where we are assuming $\omega(t),\alpha(t),\beta(t)\in%
\mathbb{C}
$. Evidently, $H\left(  t\right)  $ is not Hermitian when $\omega(t)\notin%
\mathbb{R}
$ or $\alpha(t)\neq\beta^{\ast}(t)$, and it becomes $\mathcal{PT}$-symmetric
when demanding $\omega(t)$ to be an even function in $t$ or a generic function
of $it$, simultaneously with demanding $\alpha(t),\beta(t)$ to be odd
functions in $t$ or pure-imaginary generic functions of $it$.

In order to determine the Dyson map given by Eq. (\ref{eq8}) we first consider
the same ansatz as that in Ref. \cite{Fring} for $\eta\left(  t_{0}\right)
=\exp\left[  \gamma\left(  t_{0}\right)  a+\lambda\left(  t_{0}\right)
a^{\dagger}\right]  $; we then compute the time-independent parameters
$\gamma$ and $\lambda$ from the quasi-Hermiticity relation (\ref{Eq5}) instead
of the similarity transformation (\ref{eq}), avoiding the need for the
perturbation expansion of the time-ordering operator in the TD Dyson map
$\eta\left(  t\right)  $. The relation $H^{\dag}(t)=\rho(t_{0})H(t)\rho
^{-1}(t_{0})$, coming from Eq. (\ref{Eq5}), thus demands the functions
$\omega(t)$ and $\alpha\left(  t\right)  \beta\left(  t\right)  \ $to be real
and $\gamma\left(  t_{0}\right)  +\lambda^{\ast}\left(  t_{0}\right)
=\kappa\left[  \beta^{\ast}\left(  t\right)  -\alpha\left(  t\right)  \right]
/\omega\left(  t\right)  $, such that $\left[  \gamma^{\ast}\left(
t_{0}\right)  +\lambda\left(  t_{0}\right)  \right]  \alpha\left(  t\right)
\in%
\mathbb{R}
$. With the TD functions delimited in this way and guaranteeing the
Hermiticity of $h\left(  t\right)  $, we then use the similarity
transformation (\ref{eq}) to compute
\begin{equation}
h\left(  t\right)  =2\left[  \omega\left(  t\right)  a^{^{\dagger}}a+u\left(
t\right)  a+u^{\ast}\left(  t\right)  a^{^{\dagger}}+f\left(  t\right)
\right]  , \label{eq10}%
\end{equation}
and\textbf{ }$\tilde{V}\left(  \tau\right)  =\alpha\left(  t\right)
e^{i\chi\left(  t\right)  }a+\beta\left(  t\right)  e^{-i\chi\left(  t\right)
}a^{\dag}$, with\textbf{ }$\chi\left(  t\right)  =\int_{t_{0}}^{t}%
\omega\left(  \tau\right)  d\tau$\textbf{. }Evidently, the similarity
transformation, and consequently the TD Dyson map, is as important to the
problem as the quasi-Hermiticity relation, and so the time-independent metric
operator\textbf{. }Considering the perturbation parameter $\kappa\ll1$, we
have also verified in SM (up to first order of perturbation to avoid extending
the already lengthy calculations), that $\rho(t)=\eta^{\dag}\left(
t_{0}\right)  \eta\left(  t_{0}\right)  $, now without directly using Eq.
(\ref{Eq5}), but using instead the restrictions imposed by this equation on
the TD parameters of the Hamiltonian (\ref{eq9}). Moreover, we compute the TD
functions $u\left(  t\right)  =\omega\left(  t\right)  \left[  \gamma\left(
t_{0}\right)  -i\kappa\tilde{\alpha}\left(  t\right)  \right]  +\kappa
\alpha\left(  t\right)  e^{i\chi\left(  t\right)  }$ and $f\left(  t\right)
=\left\vert u\left(  t\right)  \right\vert ^{2}/\omega(t)$, where
$\tilde{\alpha}\left(  t\right)  =\int_{t_{0}}^{t}d\tau\alpha(\tau
)e^{i\chi\left(  \tau\right)  }$.

\textit{Solutions of the Schr\"{o}dinger equation for the quasi-Hermitian
Hamiltonian}. Using the Lewis and Riesenfeld invariants \cite{LR}, as done in
Ref. \cite{PL}, the basis state solutions of the Schr\"{o}dinger equation
governed by Hamiltonian $h\left(  t\right)  $ are given by the TD displaced
number states
\begin{equation}
\left\vert \phi_{m}(t)\right\rangle =e^{i\Phi_{m}\left(  t\right)  }D\left[
\theta\left(  t\right)  \right]  \left\vert m\right\rangle ;\qquad
m=0,1,2,\ldots, \label{eq23}%
\end{equation}
where $\theta\left(  t\right)  $ follows from the equation $i\dot{\theta
}\left(  t\right)  =2\omega\left(  t\right)  \theta\left(  t\right)  +u^{\ast
}\left(  t\right)  $, whereas the TD Lewis and Riesenfeld phases are given by
\begin{equation}
\Phi_{m}\left(  t\right)  =-\int_{t_{0}}^{t}d\tau\left\{  m\varpi\left(
\tau\right)  +f\left(  \tau\right)  +\operatorname{Re}\left[  u(\tau
)\theta\left(  \tau\right)  \right]  \right\}  . \label{24}%
\end{equation}
It thus follows that $\left\vert \phi_{m}(t)\right\rangle =V(t,t_{0}%
)\left\vert m\right\rangle $, with the evolution operator $V(t,t_{0}%
)=\Upsilon(t)D\left[  \theta\left(  t\right)  \right]  R\left[  \chi\left(
t\right)  \right]  $, the rotation $R\left[  \chi\left(  t\right)  \right]
=\exp\left[  -i2\chi\left(  t\right)  a^{\dagger}a\right]  $, and the overall
phase $\Upsilon(t)=\exp\left(  -i\int_{t_{0}}^{t}d\tau\left\{  f\left(
\tau\right)  +\operatorname{Re}\left[  u(\tau)\theta\left(  \tau\right)
\right]  \right\}  \right)  $. Consequently, $\left\vert \psi_{m}%
(t)\right\rangle =\eta^{-1}(t)\left\vert \phi_{m}(t)\right\rangle $ and for a
generic superposition $\left\vert \phi(t)\right\rangle =\sum\nolimits_{m}%
c_{m}\left\vert \phi_{m}(t)\right\rangle $, the generic solution of the
Schr\"{o}dinger equation for the quasi-Hermitian $H\left(  t\right)  $ is
given by%
\begin{equation}
\left\vert \psi(t)\right\rangle =\eta^{-1}(t)\left\vert \phi(t)\right\rangle
=\eta^{-1}(t)U(t,t_{0})\left\vert \phi(t_{0})\right\rangle , \label{25}%
\end{equation}
with%
\begin{equation}
U(t,t_{0})=V(t,t_{0})V^{\dag}(t_{0},t_{0})=\Upsilon(t)D\left[  \theta\left(
t\right)  \right]  R\left[  \chi\left(  t\right)  \right]  \left(  D\left[
\theta\left(  t_{0}\right)  \right]  \right)  ^{-1}\text{.} \label{eq26}%
\end{equation}

\textit{Observables}. The observables associated with the pseudo-Hermitian
$H(t)$, given by Eq. (\ref{eq4}), are easily computed for the quadratures
$x_{\ell}=\left[  a^{\dagger}-\left(  -1\right)  ^{\ell}a\right]  /2\left(
i\right)  ^{\ell-1}$, $\ell=1,2$, leading to the operators%
\begin{equation}
\left(
\begin{array}
[c]{c}%
X_{1}\\
X_{2}%
\end{array}
\right)  =\left(
\begin{array}
[c]{cc}%
\cos\left[  \chi\left(  t\right)  \right]  & -\sin\left[  \chi\left(
t\right)  \right] \\
\sin\left[  \chi\left(  t\right)  \right]  & \cos\left[  \chi\left(  t\right)
\right]
\end{array}
\right)  \left(
\begin{array}
[c]{c}%
x_{1}\\
x_{2}%
\end{array}
\right)  +\frac{1}{2}\left(
\begin{array}
[c]{c}%
i\kappa\left[  \tilde{\alpha}\left(  t\right)  -\tilde{\beta}\left(  t\right)
\right]  -\gamma\left(  t_{0}\right)  +\lambda\left(  t_{0}\right) \\
\kappa\left[  \tilde{\alpha}\left(  t\right)  +\tilde{\beta}\left(  t\right)
\right]  +i\left[  \gamma\left(  t_{0}\right)  +\lambda\left(  t_{0}\right)
\right]
\end{array}
\right)  \label{eq28}%
\end{equation}
where the first term on the \textit{rhs} stands for the unperturbed diagonal
Hamiltonian $H_{0}\left(  t\right)  $ whereas the second term stands for the
perturbation correction. Regarding the Hamiltonian $H(t)$ itself, its matrix
elements in Fock space states are given by%
\begin{align}
\left\langle \psi_{m}(t)\right\vert H(t)\left\vert \psi_{n}(t)\right\rangle
_{\rho(t_{0})}  &  =\left\langle \phi_{m}(t)\right\vert \frac{h(t)}%
{2}\left\vert \phi_{n}(t)\right\rangle =\left\langle m\right\vert V^{\dag
}(t,t_{0})\frac{h(t)}{2}V(t,t_{0})\left\vert n\right\rangle \nonumber\\
&  =\left(  \mathcal{A}(t)\delta_{mn}+\sqrt{n+1}\mathcal{B}(t)\delta
_{m,n+1}+\sqrt{n}\mathcal{B}^{\ast}(t)\delta_{m,n-1}\right)  e^{i2\chi(m-n)}
\label{eq29}%
\end{align}
where $\mathcal{A}(t)=\omega(t)\left[  n+|\theta(t)|^{2}\right]
+2\operatorname{Re}\left[  u(t)\theta(t)\right]  +f(t)$ and $\mathcal{B}%
(t)=\omega(t)\theta(t)+u^{\ast}(t)$.

$\mathcal{PT}$\textit{-symmetry breaking}. In spite of the time-dependence of
the Hermitian Hamiltonian (\ref{eq10}), we successfully derive an eigenvalue
equation for this operator by defining, as in Ref. \cite{PL}, the TD operators
$b(t)=a+\xi^{\ast}\left(  t\right)  $ and $b^{\dag}(t)=a^{\dag}+\xi\left(
t\right)  $, associated with the relations $b^{\dagger}b|\zeta_{m}%
(t)\rangle=m|\zeta_{m}(t)\rangle$, $b|\zeta_{m}(t)\rangle=\sqrt{m}|\zeta
_{m-1}(t)\rangle$, and $b^{\dagger}|\zeta_{m}(t)\rangle=\sqrt{m+1}|\zeta
_{m+1}(t)\rangle$, where the wave vector $|\zeta_{m}(t)\rangle$ $=D\left[
-\xi^{\ast}\left(  t\right)  \right]  |m\rangle$ stands for the displaced Fock
states with $\xi\left(  t\right)  =u\left(  t\right)  /\omega(t)$. Now, up to
sencond order of perturbation, in order to allow us to analyze the
$\mathcal{PT}$-symmetry breaking, the operators $b(t)$and $b^{\dag}(t)$ help
us to rewritte Eq. (\ref{eq10}) ---with unchanged $u(t)$ but $f\left(
t\right)  =\left[  \left\vert u\left(  t\right)  \right\vert ^{2}-\kappa
^{2}\alpha\left(  t\right)  \beta\left(  t\right)  \right]  /\omega(t)$--- in
the form $h(t)=2\omega(t)b^{\dagger}b-2\kappa^{2}\alpha(t)\beta(t)/\omega(t)$,
thus leading to the TD eigenvalue equation
\begin{equation}
h(t)|\zeta_{m}(t)\rangle=\mathcal{E}_{m}(t)|\zeta_{m}(t)\rangle\text{,}
\label{eq30}%
\end{equation}
with $\mathcal{E}_{m}(t)=2\omega(t)m-2\kappa^{2}\alpha(t)\beta(t)/\omega(t)$.
From Eq. (\ref{eq30}) and the similarity transformation $h(t)=2\eta
(t)H(t)\eta^{-1}(t)$ we obtain (apart from an irrelevant factor 2)%
\begin{equation}
H(t)\left[  \eta^{-1}(t)\left\vert \zeta_{m}(t)\right\rangle \right]
=\mathcal{E}_{m}(t)\left[  \eta^{-1}(t)\left\vert \zeta_{m}(t)\right\rangle
\right]  \text{,} \label{31}%
\end{equation}
showing ---as usual in the case of time-independent non-Hermitians
Hamiltonians and Dyson maps--- that the quasi-Hermitian $H(t)$ and its
Hermitian counterpart, are isospectral partners. From the eigenvalue Eq.
(\ref{31}) its is clear that the $\mathcal{PT}$-symmetry breaking occurs if
$\omega(t)$ and/or $\alpha(t)\beta(t)$ cease to be real, resulting in the loss
of the Hermiticity of $h(t)$ \cite{Grounds}.

The eigenstates and the solutions of the Schr\"{o}dinger equation for
$h\left(  t\right)  $ are connected through the relation $\left\vert \phi
_{m}(t)\right\rangle =\mathcal{U}(t,t_{0})|\zeta_{m}(t)\rangle$ with
$\mathcal{U}(t,t_{0})=V(t,t_{0})D^{\dag}\left[  -\xi^{\ast}\left(  t\right)
\right]  $ and it is not difficult to find that the eigenstates of $h(t)$ are
the solutions of the Schr\"{o}dinger equation governed by the Hamiltonian
$\mathcal{H}(t)=\mathcal{U}^{\dag}(t,t_{0})h(t)\mathcal{U}(t,t_{0}%
)+i\hbar\mathcal{U}^{\dag}(t,t_{0})\partial_{t}\mathcal{U}(t,t_{0})$.

\textit{On the generality of the Schr\"{o}dinger-like equation}. It is worth
stressing that the Schr\"{o}dinger-like equation (\ref{eq5}) can be taken as a
general procedure for the derivation of Dyson maps, even in the case of
time-independent non-Hermitian Hamiltonians. In this case, all the
expressions, from Eq. (\ref{eq1}) to (\ref{Eq7}), remain valid except that
$H(t)$ must be replaced by $H$ and the time ordering operator must be removed
from Eq. (\ref{eq6}), thus leading to $\eta\left(  t\right)  =\eta\left(
t_{0}\right)  \exp\left[  -iH\left(  t-t_{0}\right)  \right]  $ and,
consequently, to a time-independent hermitian $h=2\eta\left(  t\right)
H\eta^{-1}\left(  t\right)  =2\eta\left(  t_{0}\right)  H\eta^{-1}\left(
t_{0}\right)  $. Therefore, for a time-independent $H$ we simply recover the
time-independent scenario for non-Hermitian quantum mechanics, showing that
the Schr\"{o}dinger-like equation can be used as a general procedure for the
derivation of TD Dyson maps with associated time-independent metric operators,
thus ensuring simultaneously the unitarity of the time evolution and the
observability of a quasi-Hermitian Hamiltonian.

Even more generally, the strategy for the derivation of the Dyson map is not
limited to the non-Hermitian quantum mechanics; it can be used when two
Hermitian Hamiltonians are connected through a unitary transformation (instead
of the non-unitary Dyson map) in the standard form $\tilde{h}(t)=\tilde
{U}(t)\tilde{H}(t)\tilde{U}^{\dag}(t)+i\left[  \partial_{t}\tilde
{U}(t)\right]  \tilde{U}^{\dag}(t)$. By defining the Schr\"{o}dinger-like
equation $i\partial_{t}\tilde{U}(t)=\tilde{U}(t)H\left(  t\right)  $, leading
to the solution $\tilde{U}(t)=\tilde{U}\left(  t_{0}\right)  T\exp\left[
-i\int_{t_{0}}^{t}d\tau\tilde{H}\left(  \tau\right)  \right]  $, the relation
between the Hamiltonians reduces to $\tilde{h}(t)=2\tilde{U}(t)\tilde
{H}(t)\tilde{U}^{\dag}(t)$, thus simplifying the form of $\tilde{h}(t)$ by
circumventing the need for a (not always easy to derive) Gauss decomposition
for the time derivative of the operator $\tilde{U}(t)$.

\textit{Conclusion}. As already stressed above, working in a scenario where TD
metric operators are considered, in Ref. \cite{Fring} it has been demonstrated
that the TD Dyson equation and quasi-Hermiticity relation can be solved
consistently at the cost of rendering the non-Hermitian Hamiltonian to be a
nonobservable quantity. Therefore, in complete analogy to the time-independent
scenario, where a time-independent Dyson map is used, it follows from Ref.
\cite{Fring} that any observable $o(t)$ in the Hermitian system possesses a
counterpart $O(t)$ in the non-Hermitian system, given by Eq. (\ref{eq4}), even
though the Hamiltonian is not an observable.

Here, disconnecting for the first time in the Literature the time-dependence
of the Dyson map from that of the metric operator, we first construct a
Schr\"{o}dinger-like equation, governed by the non-Hermitian Hamiltonian
itself, from which we derive a TD Dyson map which remarkably leads to a
time-independent metric operator. Whereas the time-independence of the metric
operator ensures the quasi-Hermiticity relation and then the unitarity of the
time evolution simultaneously with the observability of a quasi-Hermitian
Hamiltonian, the time-dependence of the Dyson map is an important demand since
for a TD non-Hermitian Hamiltonian, a time-independent Dyson map is a rather
restrictive choice.

We have shown that our Schr\"{o}dinger-like equation applies for the
derivation of a TD Dyson map either from a TD or a time-independent
non-Hermitian Hamiltonian, in the latter case recovering exactly the standard
procedure for time-independent non-Hermitian quantum mechanics. We have, in
addition, presented an illustrative example given by the harmonic oscillator
with a TD frequency under a TD non-Hermitian linear amplification process.
This Hamiltonian has been solved using the Lewis and Riesenfed TD invariants,
in a similar fashion to what has been done in Ref. \cite{Fring}, but now on a
framework where the quasi-Hermitian Hamiltonian is also an observable
quantity. We also succeeded in achieving a TD eigenvalue equation for our
quasi-Hermitian Hamiltonian, which has helped us to analyze the $\mathcal{PT}%
$-symmetry breaking process.

\begin{flushleft}
{\Large \textbf{Acknowledgements}}
\end{flushleft}

FSL would like to thank CNPq (Brazil) for support, and MHYM would like to
thank CAPES (Brazil) for support and the City University London for kind hospitality.


\begin{thebibliography}{99}                                                                                               %


\bibitem {BB}C. M. Bender and S. Boettcher, Phys. Rev. Lett. \textbf{80,} 5243 (1998).

\bibitem {Mostafazadeh}A. Mostafazadeh, J. Math. Phys. \textbf{43,} 205 (2002).

\bibitem {Grounds}C. M. Bender, Rep. Prog. Phys. \textbf{70}, 947 (2007); A.
Mostafazadeh, Int. J. Geom. Meth. Mod. Phys. \textbf{07}, 1191 (2010); C.
Bender, A. Fring, U. Gunther, and H. Jones, J. Phys. A \textbf{45}, 440301
(2012); M. Znojil, Int. J. Theor. Phys. \textbf{54}, 3867 (2015).

\bibitem {WG}P. A. Kalozoumis, C. V. Morfonios, F. K. Diakonos, and P.
Schmelcher; Phys. Rev. A \textbf{93}, 063831 (2016); H. Benisty, A. Lupu, and
A. Degiron, Phys. Rev. A \textbf{91}, 053825 (2015); V. A. Vysloukh and Y. V.
Kartashov, Optics Letters \textbf{39}, 5933 (2014); I. V. Barashenkov, L.
Baker, and N. V. Alexeeva, Phys. Rev. A \textbf{87}, 033819 (2013).

\bibitem {OL}M. Kreibich, J. Main, H. Cartarius, and G. Wunner, Phys. Rev. A
\textbf{93}, 023624 (2016); J. T. Cole, K. G. Makris, Z. H. Musslimani, D. N.
Christodoulides, and S. Rotter, Phys. Rev. A \textbf{93}, 013803 (2016); X.
Wang and J.-H. Wu, Optics Express \textbf{24}, 4289 (2016); M.-A. Miri, A.
Regensburger, U. Peschel, and D. N. Christodoulides, Phys. Rev. A \textbf{86},
023807 (2012).

\bibitem {OM}X.-W. Xu, Y.-x. Liu, C.-P. Sun, and Y. Li, Phys. Rev. A
\textbf{92}, 013852 (2015); H. Jing, S.\thinspace K. \"{O}zdemir, X.-Y.
L\"{u}, J. Zhang, L. Yang, and F. Nori, Phys. Rev. Lett. \textbf{113}, 053604 (2014).

\bibitem {Disorder}H. Vemuri, V. Vavilala, T. Bhamidipati, and Y. N. Joglekar,
Phys. Rev. A \textbf{84}, 043826 (2011); C. Mej\'{\i}a-Cort\'{e}s and M. I.
Molina, Phys. Rev. A \textbf{91}, 033815 (2015).

\bibitem {Localization}Y. V. Kartashov, C. Hang, V. V. Konotop, V. A.
Vysloukh, G. Huang, and L. Torner , Laser \& Photonics Reviews \textbf{10,
}100 (2016)

\bibitem {Chaos}C. T. West, T. Kottos, and T. Prosen, Phys. Rev. Lett.
\textbf{104}, 054102 (2010); X.-Y. L\"{u}, H. Jing, J.-Y. Ma, and Y. Wu, Phys.
Rev. Lett. \textbf{114}, 253601 (2015).

\bibitem {Solitons}T. S. Raju, T. A. Hegde, and C. N. Kumar, Journal of the
Optical Society of America B \textbf{33}, 35 (2016); M. Wimmer, A.
Regensburger, M.-A. Miri, C. Bersch, D. N. Christodoulides, and U. Peschel,
Nature Communications \textbf{6}, 7782 (2015); Y. V. Kartashov, B. A. Malomed,
and L. Torner, Optics Letters \textbf{39}, 5641 (2014); V. Achilleos, P. G.
Kevrekidis, D. J. Frantzeskakis, and R. Carretero-Gonz\'{a}lez, Phys. Rev. A
\textbf{86}, 013808 (2012).

\bibitem {Mostafa}A. Mostafazadeh, Phys. Lett. B \textbf{650}, 208 (2007);
\textit{ibid.}, arXiv:0711.0137 (2007); \textit{ibid.}, arXiv:0711.1078 (2007).

\bibitem {Controversy}M. Znojil, arXiv:0710.5653 (2007); \textit{ibid}.,
arXiv:0711.0514 (2007); ibid., Phys. Rev. D \textbf{78}, 085003 (2008); J.
Gong and Q.-h. Wang, J. Phys. A: Math. Theor. \textbf{46}, 485302 (2013); M.
Maamache, Phys. Rev. A \textbf{92}, 032106 (2015).

\bibitem {Fring}A. Fring and M. H. Y. Moussa, Phys. Rev. A \textbf{93}, 042114 (2016).

\bibitem {Swanson}A. Fring and M. H. Y. Moussa, arXiv:1606.04807 [quant-ph].

\bibitem {LR}H. R. Lewis, Jr. and W. B. Riesenfeld, J. Math. Phys. 10, 1458 (1969).

\bibitem {PL}R. R. Puri and S. V. Lawande, Phys. Lett. A 70, 69 (1979). See
also B. Baseia, S. S. Mizrahi, and M. H. Y. Moussa, Phys. Rev. A \textbf{46},
5885 (1992); S. S. Mizrahi, M. H. Y. Moussa, and B. Baseia, Int. J. of Mod.
Phys. B \textbf{8}, 1563 (1994).
\end{thebibliography}
\end{document}